\documentclass[article]{aa}
\usepackage{graphicx}
\usepackage[varg]{txfonts}
\usepackage{amstext}

\begin{document}

\title{Star clusters in the dwarf irregular galaxy Leo~A}
\author{R. Stonkut\.{e}\inst{1,2},  R. Naujalis\inst{1}, M. \v{C}eponis\inst{1}, A. Le\v{s}\v{c}inskait\.{e}\inst{1}, \and V. Vansevi\v{c}ius\inst{1,2}}

\institute{Center for Physical Sciences and Technology, Saul\.{e}tekio av. 3, 10257 Vilnius, Lithuania \and
Vilnius University Observatory, \v{C}iurlionio 29, 03100 Vilnius, Lithuania\\  \email{vladas.vansevicius@ff.vu.lt}}

\date{Submitted 15 April 2018 / Accepted 20 May 2019}

\authorrunning{R. Stonkut\.{e} et al.}
\titlerunning{Star clusters in the dwarf irregular galaxy Leo~A}

\abstract
{Leo~A is an isolated gas-rich dwarf irregular galaxy of low stellar mass and metallicity residing at the outskirts of the Local Group. Ages of the stellar populations in Leo~A range from $\sim$10~Myr to $\sim$10~Gyr. So far, only one star cluster has been discovered in this galaxy.}
{Our aim is to search for star cluster candidates in the Leo~A galaxy.}
{We performed photometry of star-like objects on HST ACS archive observation frames in the $F475W$ and $F814W$ passbands and studied the spatial clustering of the Leo~A stars down to the turn-off point of the old stellar populations.}
{We report the discovery of four star clusters in the Leo~A galaxy. This means that now a sample of five star clusters is known in Leo~A. Two clusters are of a young age ($\sim$20~Myr; overlapping with \ion{H}{II} regions) and low in mass ($\gtrsim$400~$\rm M_{\sun}$ and $\gtrsim$150~$\rm M_{\sun}$), the other three clusters are older ($\gtrsim$100~Myr) and also of low mass ($\gtrsim$300~$\rm M_{\sun}$). These rough estimates are made assuming the isochrones of a metallicity derived for \ion{H}{II} regions ($Z=0.0007$). Colour-magnitude diagrams of the stars residing in the circular areas of a 2.5\arcsec\ radius around the clusters and integrated aperture photometry results of the clusters are presented.}
{}

\keywords{galaxies: dwarf -- galaxies: star clusters -- galaxies: individual: Leo~A (DDO~69)}

\maketitle

\section{Introduction}

Studies of star clusters in low-metallicity environments help us to understand star formation processes in the high-redshift Universe. The issues of star cluster formation and evolution in various types of galaxies have been broadly discussed recently by \citet{Renaud2018}, where the importance of star cluster evolution in low-metallicity dwarf galaxies, resembling conditions of star formation in the early Universe, was stressed.
One of the best targets for testing various star cluster formation scenarios at an extremely low metallicity is the nearby dwarf irregular galaxy Leo~A. However, previous studies reported only one star cluster in this galaxy \citep{Stonkute2015} that was discovered in the deep Hubble Space Telescope (HST) Advanced Camera for Surveys (ACS) images \citep{Cole2007}. In this paper we report the discovery of four new star cluster candidates that we found in the same set of HST ACS images. 

Leo~A (Fig.~1) is an isolated dwarf irregular galaxy in the Local Group. It is a gas-rich \citep{Hunter2012} star system dominated by dark matter \citep{Brown2007} with a low metallicity \citep{vanZee2006, Kirby2017, RuizEscobedo2018}. The present-day low star formation activity is indicated by the few \ion{H}{II} regions, while the existence of an old stellar population is proven by the detection of RR~Lyr stars \citep{Dolphin2002, Bernard2013}. Stellar photometry performed with the HST Wide Field and Planetary Camera 2 (WFPC2) \citep{Tolstoy1998, Schulte-Ladbeck2002} revealed an unusual star formation history (SFH) in Leo~A: the galaxy is dominated by relatively young ($\lesssim$4~Gyr) stellar populations. Deep stellar photometry below the turn-off of the oldest populations performed with the HST ACS \citep{Cole2007} confirmed previous findings and established an SFH scenario of  a ``young galaxy''. The outer parts of the galaxy were studied with the Subaru Suprime-Cam by \citet{Vansevicius2004} and the HST Wide Field Camera 3 (WFC3) by \citet{Stonkute2018}, and revealed a presence of an extended (up to 10\arcmin) elliptical stellar envelope.

The basic parameters of the Leo~A galaxy, derived from the RGB star distribution \citep{Vansevicius2004}, are adopted in this study: centre coordinates of the galaxy, $\alpha=9^{\rm h} 59^{\rm m} 24^{\rm s}$, $\delta=+30\degr 44\arcmin 47\arcsec$ (J2000); an ellipticity, that is, the ratio of the semi-minor to the semi-major axis, $b/a = 0.6$; and a position angle of the major axis, ${\rm P.A.} = 114\degr$. The distance to Leo~A of 800~kpc \citep[$1\arcmin \approx 230$~pc;][]{Dolphin2002} is based on RR~Lyrae stars. The foreground Milky Way (MW) extinction estimates towards Leo~A are taken from \citet{Schlafly2011}, $A(F475W) = 0.068$ and $A(F814W) = 0.032$. The parameters of Leo~A are summarised in Table \ref{table:1}.

The structure of the paper is the following: Section\,\ref{sec:data} presents details of the archive observation data, reductions, and stellar photometry. Section\,\ref{sec:results} presents the results of star cluster detection and the determination of their parameters. Conclusions are presented in Section\,\ref{sec:conclusions}. 

\begin{table*}
\caption{Parameters of the Leo~A galaxy}
\label{table:1}     
\centering
\begin{tabular}{l l l}
\hline\hline
Parameter&Value&Reference \\
\hline
$\alpha$(J2000); $\delta$(J2000) & 9:59:24.0; +30:44:47 & \citet{Vansevicius2004}, RGB stars \\
b/a; P.A. & 0.60; 114\degr & \citet{Vansevicius2004}, RGB stars \\
($m-M$)${_0}$ & 24.51$\pm$0.12 (0.80$\pm$0.04~Mpc) & \citet{Dolphin2002}, RR Lyrae stars \\
$A(F475W)$; $A(F814W)$ & 0.068; 0.032 & \citet{Schlafly2011} \\
M$_{\rm STARS}$ & $3.3\cdot10^6$~$\rm{M_\sun}$ & \citet{Kirby2017} \\
M$_\ion{H}{I}$ & $6.9\cdot10^6$~$\rm{M_\sun}$ & \citet{Hunter2012} \\
M/L$_V$ & 20$\pm$6~$\rm{M_\sun}$/$\rm{L_\sun}$ & \citet{Brown2007}, B supergiants and \ion{H}{II} zones \\
12+$\log$(O/H) & 7.38$\pm$0.1 & \citet{vanZee2006}, a planetary nebula and \ion{H}{II} zones \\
12+$\log$(O/H) & 7.4$\pm$0.2 & \citet{RuizEscobedo2018}, \ion{H}{II} zones \\
<[Fe/H]> & $-1.67_{-0.08}^{+0.09}$ & \citet{Kirby2017}, RGB stars \\
\hline
\end{tabular}
\tablefoot
{$\alpha$(J2000) and $\delta$(J2000) are the equatorial coordinates of the galaxy centre; b/a is the ratio of minor to major axes; P.A. is the position angle of the major axis; ($m-M$)${_0}$ is the true distance modulus; $A(F475W)$ and $A(F814W)$ are the foreground extinction in corresponding passbands; M$_{\rm STARS}$ is the stellar mass; M$_\ion{H}{I}$ is the mass of neutral hydrogen; M/L$_V$ is the lower limit of the mass-to-luminosity ratio; 12+$\log$(O/H) is the oxygen abundance; and <[Fe/H]> is the average metallicity.}
\end{table*}

\section{Observation data and stellar photometry}
\label{sec:data}

\begin{figure*}
\centering
\includegraphics[scale=0.65]{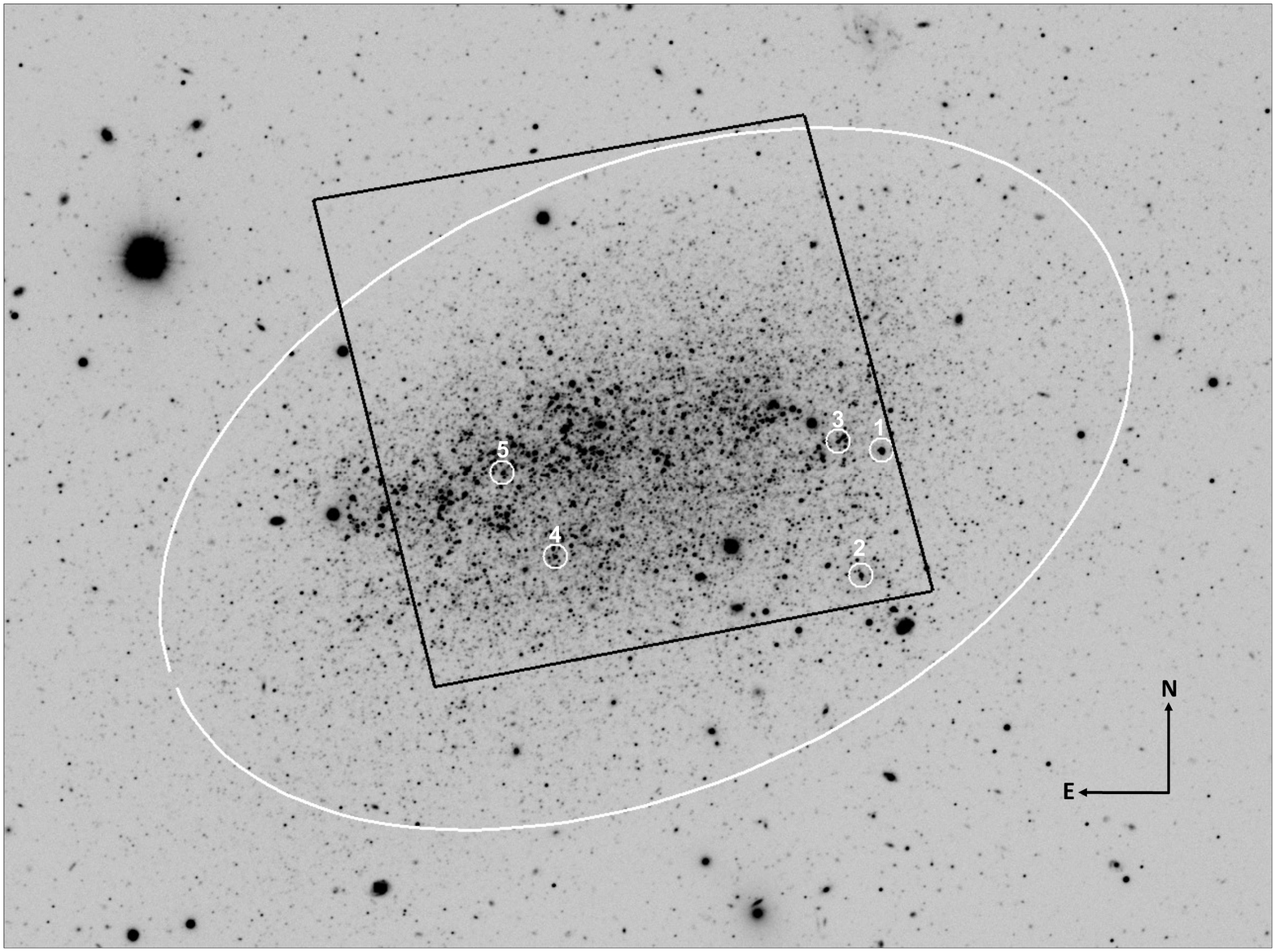}
\caption{\small Subaru Suprime-Cam $B$-passband image of the Leo~A galaxy \citep{Stonkute2014}. The ellipse ($b/a = 0.6$; ${\rm P.A.} = 114\degr$) of the semi-major axis equal to the Holmberg radius, $a = 3.5\arcmin$ \citep{Mateo1998}, centred at $\alpha=9^{\rm h} 59^{\rm m} 24^{\rm s}$, $\delta=+30\degr 44\arcmin 47\arcsec$ (J2000), is shown by the white line. The HST ACS field \citep{Cole2007} is marked by the black line. Discovered star clusters are marked by white open circles. North is up, east is left.}
\label{fig:fig1}
\end{figure*}

Stellar photometry and integrated photometry of star clusters were performed using HST ACS frames in the $F475W$ and $F814W$ passbands from the project Local Cosmology from Isolated Dwarfs (LCID; http://www.iac.es/proyecto/LCID), see \citet{Cole2007} for observation details. 

The archival data were downloaded from the Mikulski Archive for Space Telescopes (MAST). We retrieved bias-subtracted, flat-fielded, charge transfer efficiency (CTE) corrected HST ACS {\tt flc} images produced by the STScI ``on-the-fly reprocessing'' (OTFR) pipeline OPUS versions 2015\_2b, which used CALACS version 8.3.0.

To perform stellar photometry we used the software package DOLPHOT 2.0 \citep[][and many unpublished updates]{Dolphin2000}. We followed the recommended preprocessing steps and the photometry recipe provided in the manual for the HST ACS module (version of 24 February 2016). We used {\tt AstroDrizzle 1.1.16} (default parameter values) to create clean, deep-drizzled reference frames for object detection and coordinate transformations from 16 sub-exposures in each of the $F475W$ and $F814W$ passbands. This also allowed us to flag cosmic rays in the individual {\tt flc} images and to update data quality images.

We used the values of the DOLPHOT parameters recommended in the HST ACS manual: the {\tt FitSky} parameter was set to 1, which means the sky fitting in an annulus around each star ($\rm{R}_{inner} = 15$, $\rm{R}_{outer} = 35$ pixels) and the point-spread function (PSF) fitting inside a radius of $\rm{R}_{apert} = 4$ pixels. 

DOLPHOT determines magnitudes, magnitude errors, object fit, and shape parameters in individual {\tt flc} frames, and then combines them per filter. To combine the magnitudes we set a parameter {\tt FlagMask = 5}, which means that only measurements with error flags equal to 0 (excellent photometric quality) and 2 (bad or saturated pixels are present) were used. 

In order to optimise the parameters for the photometry, we performed numerous tests with various source detection thresholds ({\tt SigFind} and {\tt SigFinal}), and a minimum allowed separation for two stars ({\tt RCombine}). The best photometry quality was achieved with {\tt SigFind = 1.5}, {\tt SigFinal = 4}, and {\tt RCombine = 1.5}. These parameter values were applied for the final photometry. We also set the parameter {\tt Force1 = 1} (all detected sources are fitted as stars), as is suggested for crowded field photometry, and for the further analysis, we selected only stars with the object type flag equal to 1 and a signal-to-noise ratio $\geq$5.0.

The initial photometry catalogue contained measurements in two passbands of 189\,947 objects. In order to clean the photometry catalogue, combined mosaic images were visually inspected, and objects falling on obvious background galaxies, bright MW stars, or image artefacts, as well as those residing closer to the frame borders than $2\times\rm{R}_{apert} = 8$ pixels, were rejected. In order to further clean the catalogue from extended objects, we rejected objects by sharpness in both passbands ($0.2<{\tt sharpness_{\it F814W}}<-0.2$ and $0.2<{\tt sharpness_{\it F457W}}<-0.2$). This left us with $\rm{N}=151\,146$ objects. 

\section{Results and discussions}
\label{sec:results}

\begin{figure*}
\centering
\includegraphics[scale=0.68]{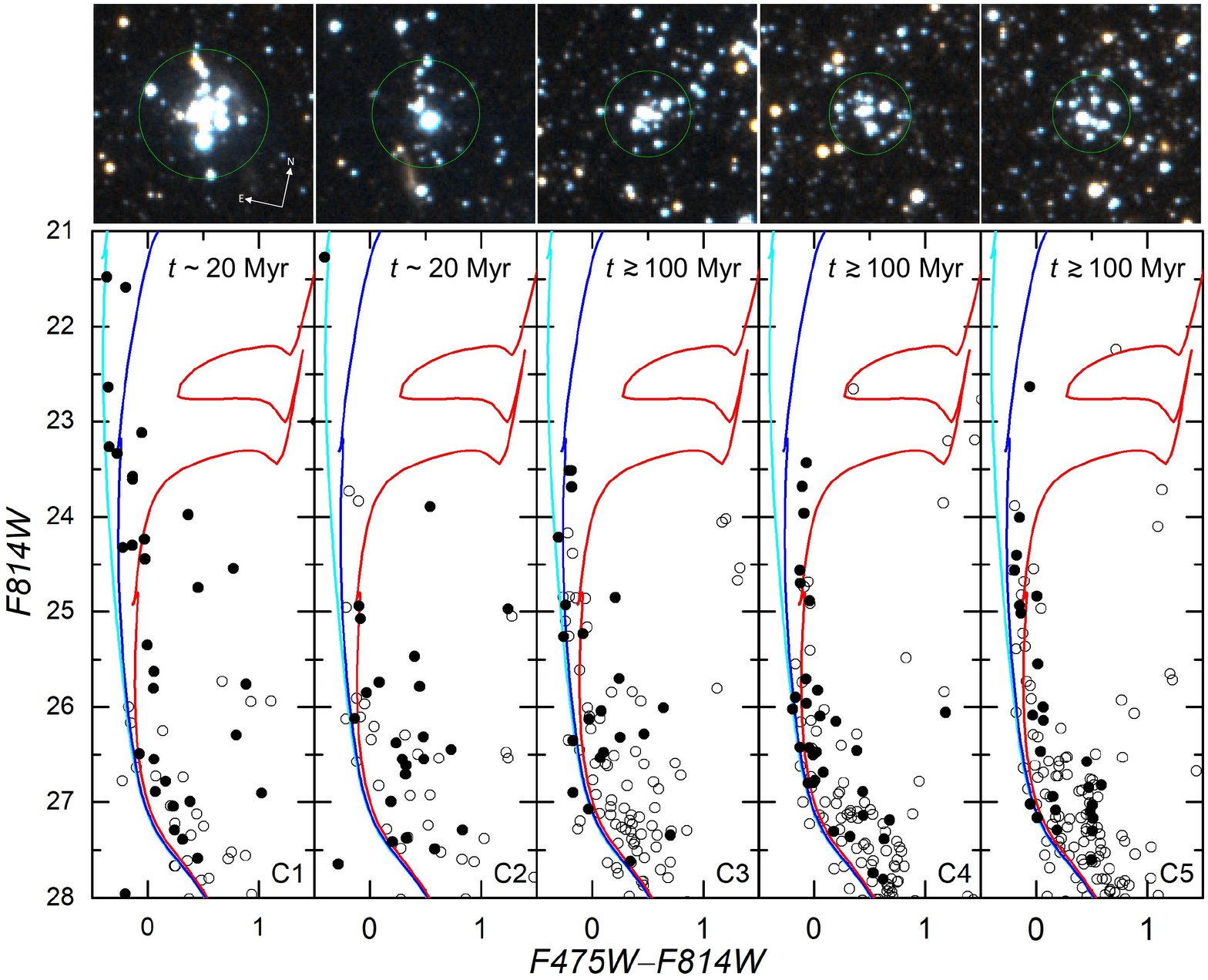}
\caption{\small Colour images ($5\arcsec\times5\arcsec$) of the clusters (identification codes are shown in lower right corners of each panel) and their CMDs showing the star-like objects residing inside the green circle (radius is equal to R, see Table \ref{table:2}) marking the cluster itself (filled black circles) and star-like objects residing inside the circle of $2.5\arcsec$ radius (open circles). The colour images of clusters are constructed from the HST ACS frames taking images in the $F814W$, ($F475W+F814W$)/2, and $F475W$ passbands as an input to RGB channels. The PARSEC isochrones, release v1.2S \citep{Bressan2012}, of $Z=0.0007$ metallicity and ages of 20~Myr (the cyan line), 100~Myr (the blue line), and 500~Myr (the red line) are shown in all panels. All isochrones are adjusted for the distance modulus of 24.51 and MW foreground reddening, $A(F475W)=0.068$ and $A(F814W)=0.032$.}
\label{fig:fig2}
\end{figure*}

\begin{table*}
\caption{Parameters of star clusters in Leo~A}
\label{table:2}     
\centering
\begin{tabular}{c c c c c c c c c c c}
\hline\hline
ID&$\alpha$(J2000)&$\delta$(J2000)&R [\arcsec]&$\rm{R_h}$ [\arcsec]&$F814W$&$F814W_{\rm 2R_h}$&$CI$&$CI_{\rm 2R_h}$&age (Myr)&mass ($\rm{M_\sun}$) \\
(1)&(2)&(3)&(4)&(5)&(6)&(7)&(8)&(9)&(10)&(11) \\
\hline
LeoA-C1 & 9:59:16.5 & +30:44:59 & 1.50 & 0.55 & 19.85 & 19.96 & -0.11  & -0.11  &  $\sim$20 &  $\gtrsim$400 \\
LeoA-C2 & 9:59:17.2 & +30:44:08 & 1.25 & 0.7$^*$ & 20.82 & 20.79 & -0.24  & -0.23  &  $\sim$20 &  $\gtrsim$150 \\
LeoA-C3 & 9:59:17.9 & +30:45:02 & 1.00 & 0.31 & 21.69 & 21.86 & -0.15  & -0.14  & $\gtrsim$100 &  $\gtrsim$300 \\
LeoA-C4 & 9:59:26.9 & +30:44:15 & 0.95 & 0.37 & 21.68 & 21.89 &  0.02  & -0.06  & $\gtrsim$100 &  $\gtrsim$300 \\
LeoA-C5 & 9:59:28.5 & +30:44:50 & 0.90 & 0.34 & 21.78 & 21.85 & -0.14  & -0.15  & $\gtrsim$100 &  $\gtrsim$300 \\
\hline
\end{tabular}
\tablefoot
{(1) ID is the cluster identification number; (2) $\alpha$(J2000) and (3) $\delta$(J2000) are the equatorial coordinates; (4) R is the radius of the cluster in arcsec (in Fig. \ref{fig:fig2} it is marked with a green circle); (5) $\rm{R_h}$ is the half-light radius of the cluster derived from the photometric growth curve in arcsec (the asterisk shows the half-light radius of the cluster LeoA-C2 estimated from the star number count); (6) $F814W$ is the magnitude measured through the aperture of radius R; (7) $F814W_{\rm 2R_h}$ is the magnitude measured through the aperture of radius $2\times \rm{R_h}$; (8) $CI$ is the colour index $F475W-F814W$ measured through the aperture of radius R; (9) $CI_{\rm 2R_h}$ is the colour index $F475W-F814W$ measured through the aperture of radius $2\times \rm{R_h}$; (10) age (Myr) is a rough estimate of the cluster age in Myr; and (11) mass ($\rm{M_\sun}$) is a rough estimate of the cluster mass in solar masses, $\rm{M_\sun}$.}   
\end{table*}

\begin{figure}
\centering
\includegraphics[scale=0.19]{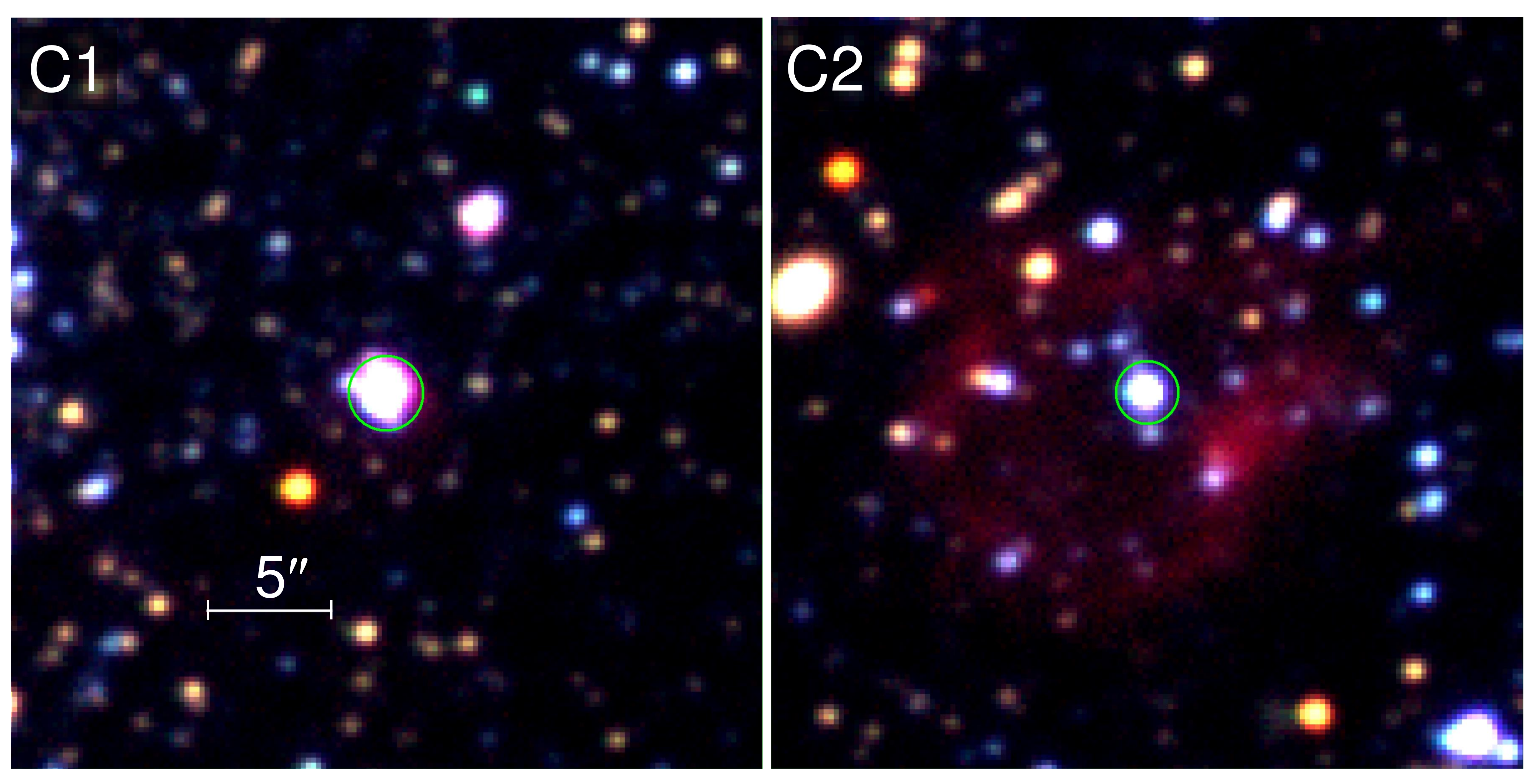}
\caption{\small Composite colour images ($30\arcsec\times30\arcsec$) of star clusters LeoA-C1 and LeoA-C2. The colour images of the clusters are made from the Subaru Suprime-Cam frames \citep{Stonkute2014} taken in the $H\alpha$, $V$, and $B$ passbands as input to the RGB channels. The green circles that mark the clusters are of the same size as in Fig. \ref{fig:fig2}. North is up, east is left.}
\label{fig:fig3}
\end{figure}

Compact star cluster recognition in galaxies at the outskirts of the Local Group based on images, taken even at observatories located in the best astroclimate conditions, is a very difficult task \citep{Narbutis2008, Stonkute2008}. To search for the compact low-mass star clusters in the Leo~A galaxy we therefore used available HST ACS archive images in the $F475W$ and $F814W$ passbands and colour-magnitude diagrams (CMDs) constructed from the stellar photometry data described in Section \ref{sec:data}. 

To search for star clusterings we selected reliably identified and measured stars ($F814W < 28$), and excluded the reddest stars ($F475W-F814W > 2.0$), which most probably are foreground MW stars. In total, we used 98\,525 stars within the HST ACS field and analysed their clustering. We used three circular cluster kernel sizes of radii 0.50\arcsec, 0.75\arcsec, and 1.00\arcsec\ in order to reveal compact ($\lesssim$2~pc) and more extended ($\sim$4~pc) star clusterings. The individual background number density for each kernel position was estimated in an annulus from 1.0\arcsec\ to 2.0\arcsec\ around the kernel centre, and then was subtracted from the number density determined within the kernel. As a result, three number density maps were produced and intercorrelated. The star number density map obtained using the kernel of 0.50\arcsec\ radius contained all significant star clusterings seen in two other maps derived with larger kernels. The further analysis was therefore based on this star number density map. 

As the first step, we calculated the ratio of star number densities within a radius of 0.5\arcsec\ and an annulus background zone from 1.0\arcsec\ to 2.0\arcsec\ radius for each pixel. Arbitrarily assuming a lower threshold for the star number density ratio equal to 5, we ended up with 76 pre-selected star clusterings worth studying in more detail. However, a careful inspection of the pre-selected candidates on the HST ACS mosaics independently by five team members, using the by-eye procedure described in detail by \citet{Johnson2012}, left us with the 5 most prominent star cluster candidates. It is worth noting, however, that the super-position of stars could only marginally affect identification reliability of cluster candidates because of well-resolved stellar populations and a comfortably oriented galaxy disk. 

The colour images of the clusters and their CMDs (Fig. \ref{fig:fig2}) show the star-like objects (filled black circles in CMDs) residing inside the green circle that marks the cluster itself and star-like objects (open circles) residing inside the circle of 2.5\arcsec\ radius. The colour images of clusters are constructed from the HST ACS frames taking images in the $F814W$, ($F475W+F814W$)/2, and $F475W$ passbands as an input to RGB channels. 

The determination of an accurate centre position of the well-resolved star cluster is a sensitive procedure. For this purpose we used peaks on the spatially smoothed star number density map at the positions of suspected star clusters. An integrated growing circular aperture (the aperture radius increases by 0.1\arcsec\ up to the radius of 2.5\arcsec) photometry around these positions was performed, and radii of the first sudden flattening of the growth curves were determined for each cluster candidate. These radii were assumed as measures of cluster sizes (the parameter R in Table \ref{table:2}). The half-light radii $\rm{R_h}$ were measured on photometric growth curves at the levels corresponding to the magnitudes derived at radii R plus 0.75. The radii of star clusters (R and $\rm{R_h}$) estimated independently from the photometric growth curves in the $F475W$ and $F814W$ passbands coincide well within half of the applied aperture growth step, 0.05\arcsec. We note, however, that for the cluster LeoA-C2, the growth curve based on the star number count was used instead because a peculiar photometric growth curve arose as a result of much fainter stars that surrounded an extremely bright star.

The integrated magnitudes and colour indices, derived through the circular apertures of radii R and $2\times R_{\rm h}$, are listed in Table \ref{table:2}. The estimated accuracy of the integrated aperture photometry (magnitudes and colour indices), taking into account the errors in background level and magnitude zero-point, is $\sim$0.05. Moreover, the correct sky background subtraction is critical for determining the shape of the photometric growth curves in the outer regions of the clusters; for a detailed discussion, see \citet{Hill2006}. In order to estimate possible errors of the sky background subtraction, we constructed the cluster photometric profiles with sky background estimates derived in annuli of  different sizes in the radii range of $1.5-3.0\arcsec$. However, this led to relatively small half-light radii changes: $\Delta\rm{R_h} \sim \pm 0.05\arcsec$.

To estimate the age of star clusters, we used PARSEC isochrones, release v1.2S \citep{Bressan2012}, with $Z=0.0007$ metallicity. Isochrones of 20~Myr (a cyan line), 100~Myr (a blue line), and 500~Myr (a red line) are shown in all panels of Fig. \ref{fig:fig2} for reference. The isochrones are adjusted assuming the Leo~A distance modulus of 24.51 \citep{Dolphin2002} and a foreground MW extinction, $A(F475W)=0.068$ and $A(F814W)=0.032$ \citep{Schlafly2011}. The decision to use isochrones of $Z=0.0007$ metallicity is based on the oxygen abundance ($12+\log(O/H)=7.4$) determined for \ion{H}{II} regions in Leo~A by \citet{vanZee2006} and \citet{RuizEscobedo2018}. We note, however, that the young ages estimated from the upper main-sequence stars of the clusters are rather insensitive to the assumed metallicity. An independent strong support of the young ages of the LeoA-C1 and LeoA-C2 clusters is their location in the areas of prominent \ion{H}{II} zones (Fig. \ref{fig:fig3}). 

Based on the ages of the star clusters, we estimated their approximate mass to be in the range of $\sim$200--500~${\rm M_\sun}$ (Table \ref{table:2}). The mass of star clusters was determined from the distribution of the blue ($F475W-F814W < 0.5$) bright ($F814W < 26$) stars located on or near the main sequence (Fig. \ref{fig:fig2}). We applied the initial mass function by \citet{Kroupa2002} in the stellar mass range from 0.08 to 120~${\rm M_\sun}$. 

In order to estimate the fraction of stars that form in clusters, we employed the recent SFH in Leo~A derived within the area of the HST ACS field \citep{Ceponis2018}. Taking into account two young clusters (LeoA-C1 and LeoA-C2) and stars formed during the last 30~Myr, we determine that $\sim$10\% of stars were formed in clusters. Taking into account all five clusters (LeoA-C1--C5) and stars formed during the last 200~Myr, we derive that $\sim$2\% of stars were formed in clusters. A large difference of these estimates arises, most probably, from selection effects (detection incompleteness of older clusters) and cluster destruction or dissolution processes.

Finally, we would like to stress that all physical parameters derived for the low-mass ($\sim$300~${\rm M_\sun}$) star clusters are subject to strong stochastic effects \citep{deMeulenaer2014} and should be treated carefully just as best-guess estimates. 

\section{Conclusions} 
\label{sec:conclusions}

The aim of this study was to search for compact star cluster candidates in the dwarf irregular galaxy Leo~A, which has an extremely low metallicity \citep{Kirby2017}. So far, only one star cluster (LeoA-C1), discovered recently by \citet{Stonkute2015}, was known in this galaxy. 

We have performed photometry of star-like objects on HST ACS archive frames in the $F475W$ and $F814W$ passbands covering the centre of Leo~A and have studied the spatial clustering of the stars down to the turn-off point of old stellar populations. We found four new compact low-mass star cluster candidates. Based on the cluster CMDs analysis applying the isochrones \citep{Bressan2012} with metallicity $Z=0.0007$, which is derived for \ion{H}{II} regions \citep{vanZee2006, RuizEscobedo2018}, relatively young cluster ages and low masses were estimated (LeoA-C2: $\sim$20~Myr and $\gtrsim$150~$\rm M_{\sun}$; LeoA-C3--C5: $\gtrsim$100~Myr and $\gtrsim$300~$\rm M_{\sun}$). 

The finding of such low-mass ($\sim$300~$\rm M_{\sun}$) and young- to intermediate-age ($\sim$20--100~Myr) star clusters in the low stellar mass ($3.3\cdot10^6$~$\rm{M_\sun}$) and the extremely low metallicity ($12+\log(O/H)=7.4$) dwarf irregular galaxy Leo A (in which CO emission has not been detected so far) could help constrain star formation scenarios in the early Universe. The problems of low-mass star clusters that reside in similar environments have been addressed only recently: an extensive study of star clusters in the LEGUS dwarf galaxies has been published by \citet{Cook2019}. The Leo~A galaxy is at the lowest mass and at the lowest metallicity limits of the LEGUS dwarf galaxies. Therefore, clusters discovered in Leo~A consistently extend the parameter space of star clusters measured in the LEGUS dwarfs.

However, in order to determine a complete census of star clusters in the Leo~A galaxy, a much larger field ($\text{about four}$ times larger), observed with the resolution of the HST ACS or WFC3 cameras, is needed. This conclusion is supported by the number of additional star cluster candidates seen in the Subaru Suprime-Cam frames, for instance, the obvious young star cluster embedded in the \ion{H}{II} region resides just outside the HST ACS field; see $\sim$8\arcsec\ to the north-west of LeoA-C1 in Fig. \ref{fig:fig3}. 

\begin{acknowledgements}
We thank the anonymous referee for helpful suggestions that improved the presentation of this paper. The research has made use of the SAOImage DS9, developed by Smithsonian Astrophysical Observatory. The data presented in this paper were obtained from the Multimission Archive at the Space Telescope Science Institute. This research was funded by a grant (No. LAT-09/2016) from the Research Council of Lithuania.
\end{acknowledgements}

\bibliographystyle{aa}
\bibliography{33236corr}
\end{document}